\documentclass[iop]{emulateapj}
\usepackage{amsmath,color}
\usepackage{natbib}

\catcode`_=\active
\newcommand_[1]{\ensuremath{\sb{\mathrm{#1}}}}

\newcommand{\revise}[1]{#1}

\usepackage{hyperref}
\usepackage[normalem]{ulem}
\hypersetup{
   colorlinks=true,
   citecolor=blue,    
   linkcolor=blue,
   filecolor=magenta,      
   urlcolor=blue,
}

\newcommand{\be}[1]{\begin{equation} \label{eq:#1}}
\newcommand{\ee}{\end{equation}}
\newcommand{\ba}[1]{\begin{eqnarray} \label{eq:#1}}
\newcommand{\ea}{\end{eqnarray}}

\newcommand{\pref}{\protect\ref}
\newcommand{\iris}{{\em IRIS}}

\newcommand{\solrad}{\ifmmode{R}_{\rm S}\else${R}_{\rm S}$\fi}
\newcommand{\solmas}{\ifmmode{M}_{\rm S}\else${M}_{\rm S}$\fi}

\newcommand{\tintu}{\ifmmode{\rm erg~cm^{-2}~s^{-1}sr^{-1}}\else 
  erg~cm$^{-2}$~s$^{-1}$~sr$^{-1}$\fi}
\newcommand{\fluxu}{\ifmmode{\rm erg~cm^{-2}~s^{-1}}\else 
  erg~cm$^{-2}$~s$^{-1}$\fi}
\newcommand{\velu}{$\,$km$\,$s$^{-1}$}

\newcommand{\wave}{\ifmmode{\lambda} \else$\lambda$\fi}

\newcommand{\hot}{\ifmmode{8\times10^4~{\rm K}}\else{$8\times10^4$~K}\fi}

\newcommand\lta { \mathrel {\hbox to 0pt {\lower 3.7pt \hbox{$\sim$}
      \hss} \raise 1.7pt \hbox{$<$}}}
\newcommand\gta { \mathrel {\hbox to 0pt {\lower 3.7pt \hbox{$\sim$}
      \hss} \raise 1.7pt \hbox{$>$}}}

\newcommand{\philemail}{judge@ucar.edu}
\newcommand{\alinaemail}{alina.donea@monash.edu}
\newcommand{\danielaemail}{daniela.lacatus@monash.edu}

\newcommand{\figsjib}{
\begin{figure}[t] 
\epsscale{1.15}
{\plotone{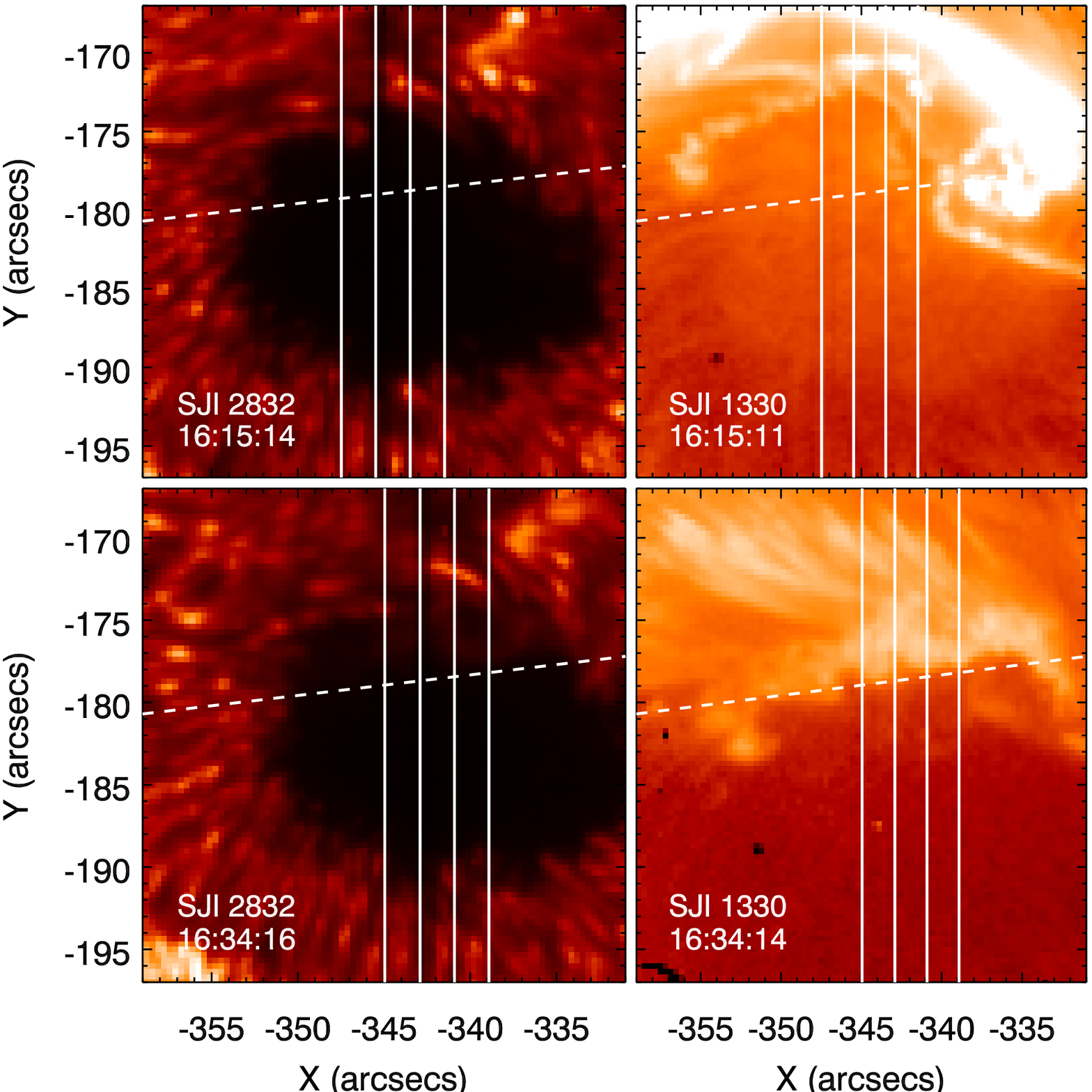}  
\vskip -5pt}
\caption{\label{fig:sjib} 
\iris{} SJI images in the 2832 channel (left) and the 1330 channel (right), before and after the impulsive phase. A logarithmic scaling has been applied to SJI 1330. The vertical lines are the positions of the \iris{} slits and the oblique dashed line is the lower boundary (in the N-S direction) of the peculiar spectra. \bf Plasma flows can be seen above this dashed line in the SJI 1330 post-flare images.}
\end{figure}
}

\newcommand{\figmgwidint}{
\begin{figure}[t] 
\epsscale{1.15}
{\plotone{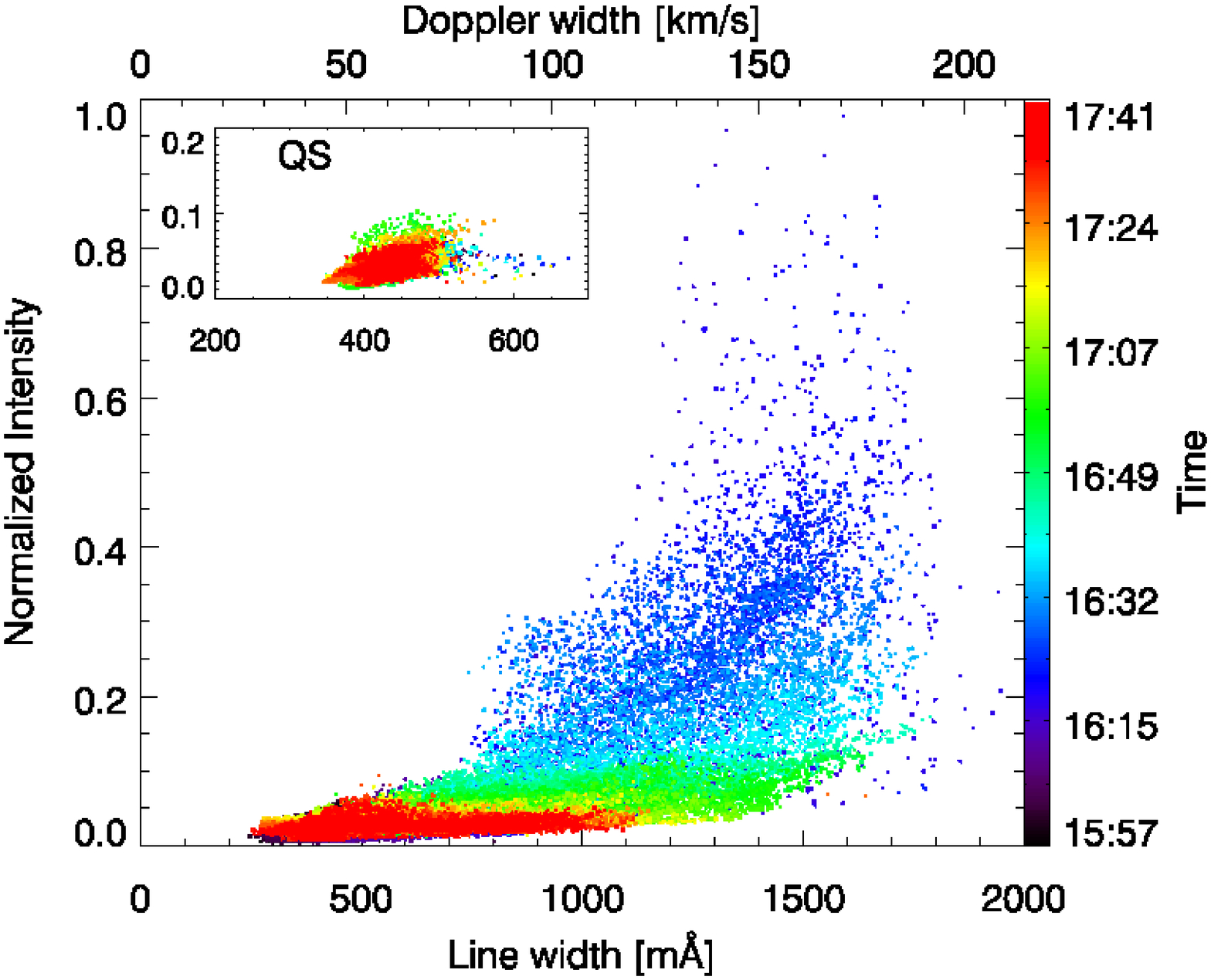}  
\vskip -5pt}
\caption{\label{fig:widint} 
\bf Normalized Intensity vs line width correlation for all the pixels between -170$\arcsec$ and -180$\arcsec$. The color represents the time of the observation and the insert represents the data from a QS patch outside the active region. }
\end{figure}
}

\newcommand{\figdopp}{
\begin{figure}[h] 
\epsscale{1.15}
{\plotone{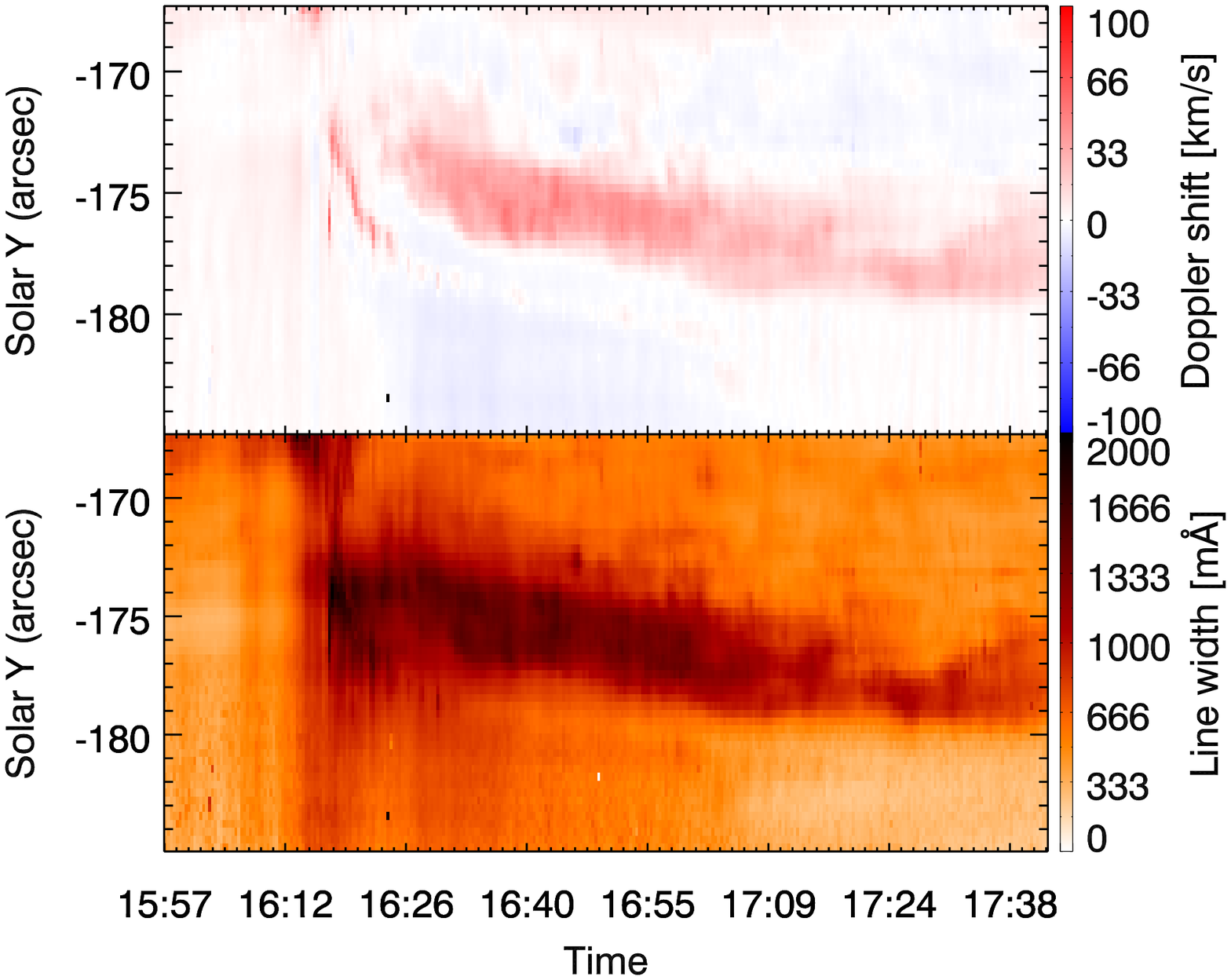}  
\vskip -5pt}
\caption{\label{fig:dopp} 
\bf Doppler shift (top) and line widths (bottom) maps for the Mg II k-line, showing the time evolution of the extended redshifted emission for the first slit position. The peculiar emission sweeps the same region as the flare ribbon. Positive Doppler shift values refer to red shift and negative values to blue shift.}
\end{figure}
}

\newcommand{\figvmgram}{
\begin{figure}[h] 
\epsscale{1.25}
{\plotone{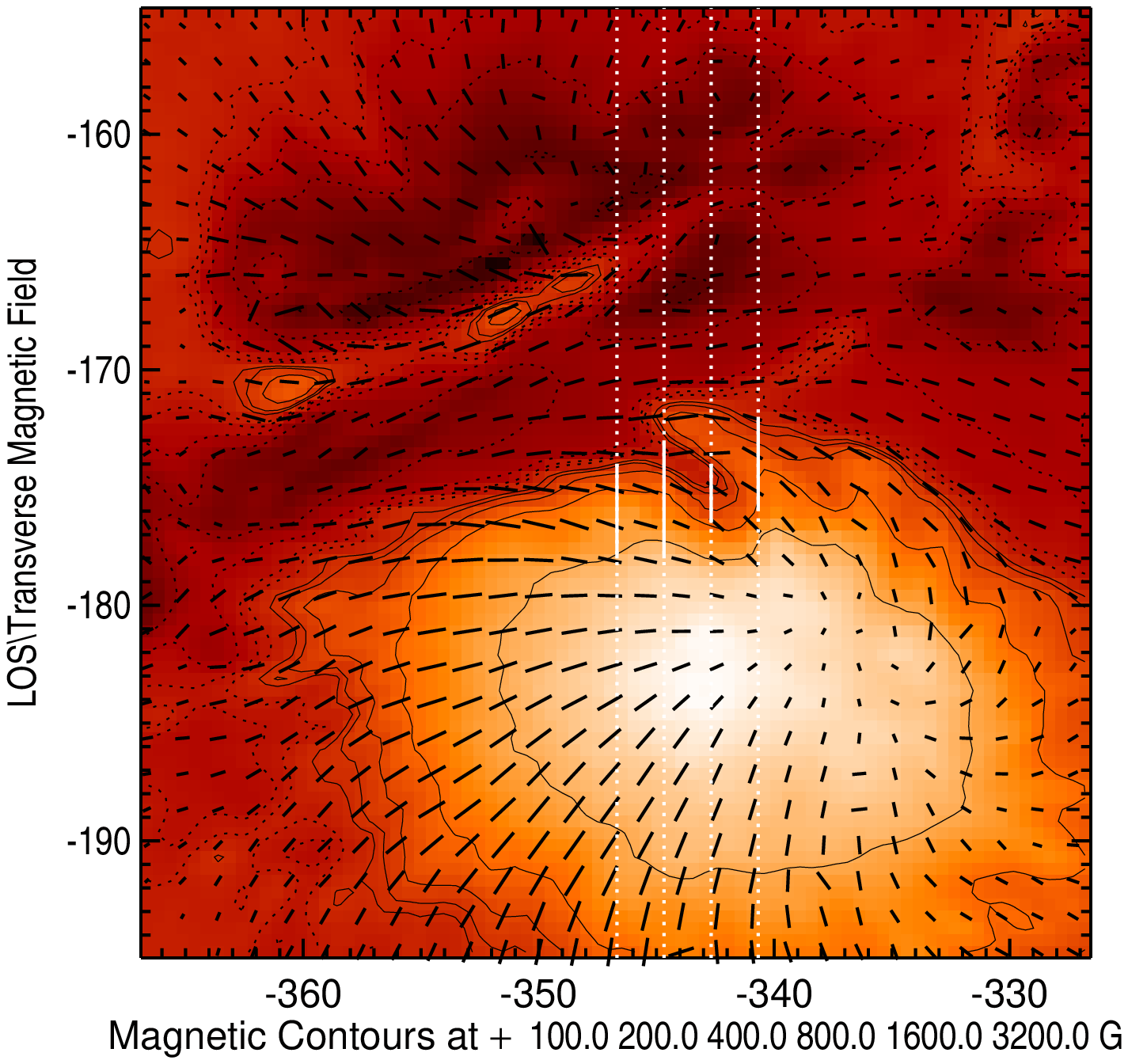}  
\vskip 12pt}
\caption{\label{fig:vmgram} 
 Vector magnetic field parameters at 16:24
UT.  The vertical dashed lines show
the positions of the IRIS slit, the solid portion of
which approximately shows the extent of very broad
lines seen in the \ion{Mg}{2} $h$ and $k$ lines shown
in previous figures.  }
\end{figure}
}

\newcommand{\figmgii}{
\begin{figure}[t] 
\epsscale{1.15}
{\plotone{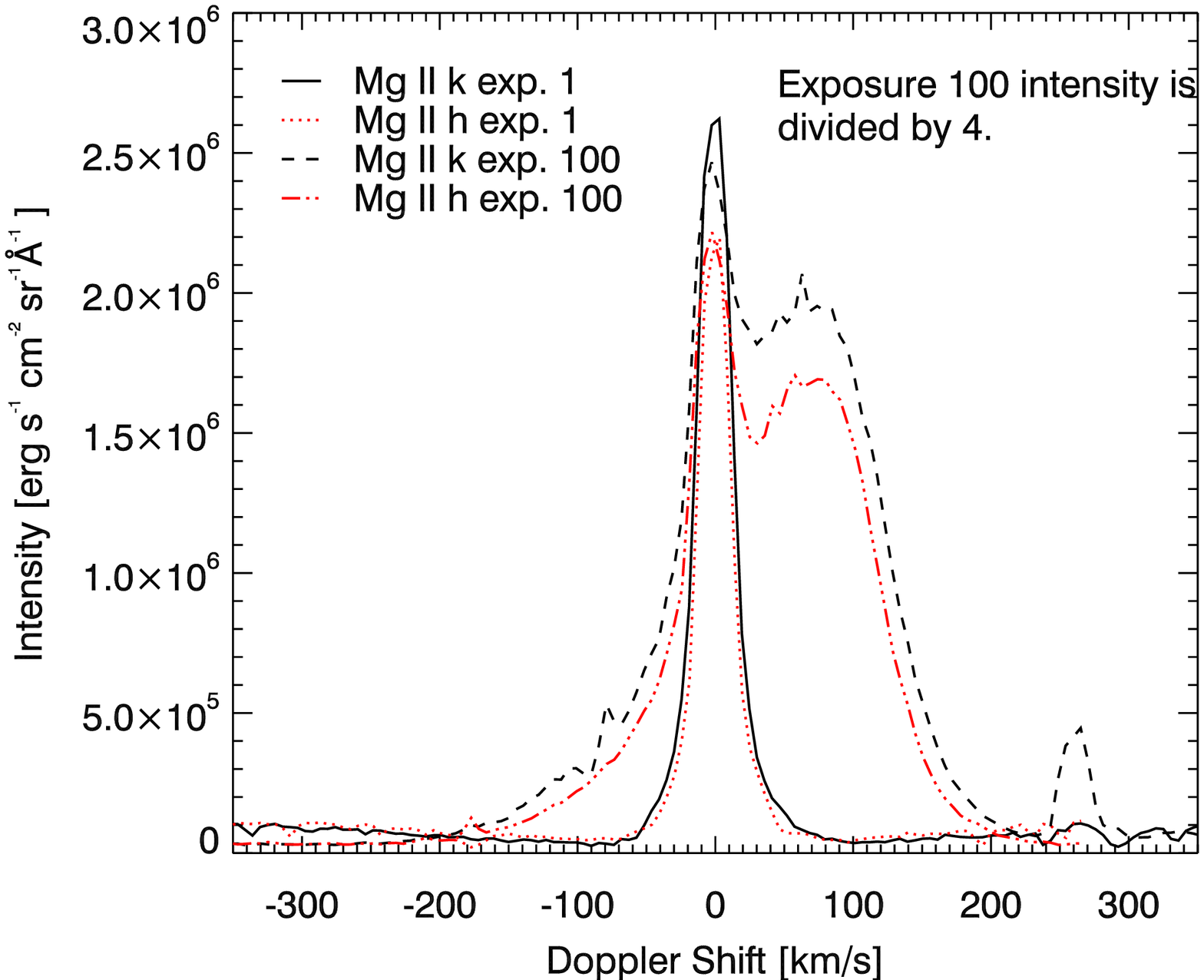}  
\vskip -5pt}
\caption{\label{fig:mgii} 
Typical intensity profiles of the \ion{Mg}{2} $h$ and $k$ lines
are shown as a function of Doppler shift, positions indicated by arrows in Figure \pref{fig:mgim}. Two profiles shown are 
those of 15:57:51 UT (exp. 1), from the pre-flare phase, and those of 
16:32:31 UT (exp. 100), from the post-flare phase.  The very broad profiles
seen during the 16:32:31 UT image  are the focus of the present work.  }
\end{figure}
}

\newcommand{\figmgim}{
\begin{figure}[h] 
\epsscale{1.15}
{\plotone{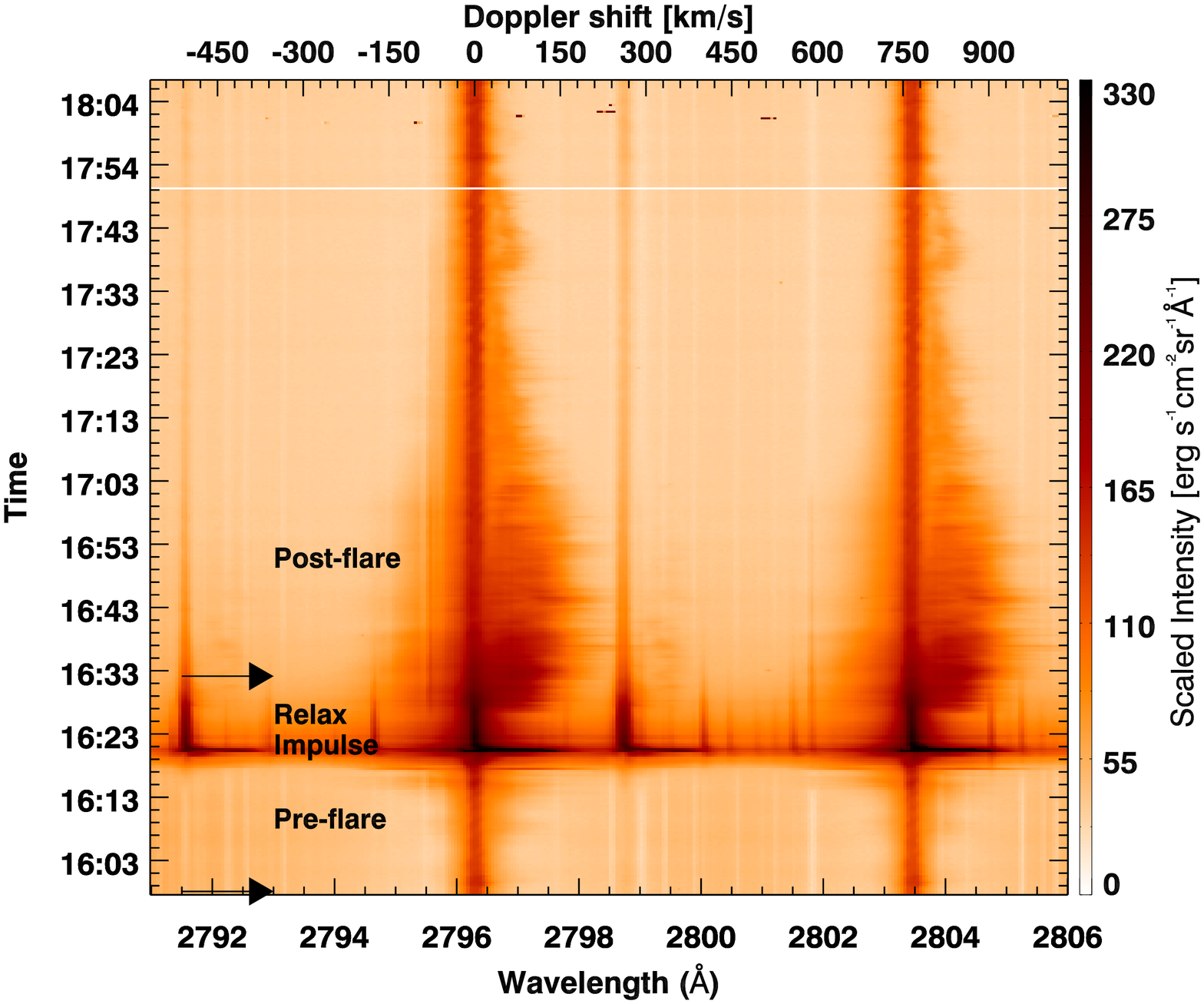}  
\vskip -5pt}
\caption{\label{fig:mgim}\bf Cubic root scaled intensity profiles of the \ion{Mg}{2} $h$
and $k$ lines (for the first slit position, at y=-175") are shown as a function of wavelength and time. Doppler shift values with respect to the k-line position are also included. The two  profiles shown in Figure \pref{fig:mgii} are from 
15:57:51 and 16:32:31 UT, indicated by arrows. }
\end{figure}
}

\newcommand\tableonenew{
\begin{table}
\label{tab:intensitiesnew}
\begin{center}
 \caption{Observed frequency-integrated emission-line intensity (\tintu)}
 \begin{tabular}{lrrrr}     
 \hline
 \hline
Multiplet & \ion{Mg}{2}  & \ion{Mg}{2} & \ion{C}{2} & \ion{Si}{4} \\
       & $h+k$  &  $3d-3p$ & 1334  & 1403 \\
Phase &         &          &   +1335  &  \\
\hline
\\
QS  &
	$5.36\times10^5$ & \ldots &	$6070$ & $ 572$  \\
\\
Pre-flare & $1.83\times10^6$ & 	\ldots & $5.16\times10^4$ &2840
\\
\\
Impulse & $4.36\times10^7$  & $\approx 1.36\times10^7$ & $1.73\times10^7$&$>3.86\times10^6$\\
\\
Relax & $2.77\times10^7$ & $4.81\times10^6$ & $4.39\times10^6$&$3.51\times10^5$\\
\\
{\bf Post-flare} &	$\mathbf{1.82\times10^7}$& $\mathbf{\approx 4.51\times 10^5}$	
	& $\mathbf{1.45\times10^6}$ &
	$\mathbf{2.04\times10^5}$\\
\hline
\end{tabular}
\end{center}
{Quiet Sun data are from the darkest regions along
the \iris{} slit. We used calibration factors returned by
iris\_get\_response.pro: dn2phot\_sg = 4.0 (FUV) and 18.0(NUV), and effective
areas of 0.474, 1.042 and 0.23 cm$^2$ for the 1335,
1400 and 2800 \AA\ wavelengths, respectively. The intensities in 
{\bf boldface} will be used throughout this article. }
\end{table}}

\newcommand\tabletwo{
\begin{table}
\label{tab:atomic}
\begin{center}
 \caption{Atomic transitions of interest}
 \begin{tabular}{llrrrr}     
 \hline
 \hline
Ion  & $\lambda$ \AA{} & upper level & lower level & gf &$\Upsilon(T_{e})$ \\
\hline
\\
\ion{Mg}{2} & 2802.705  & $3p~^2{\rm P}^o_{1/2}$ & $3s~^2{\rm S}_{1/2}$ & 0.64 & 5.6\\
            & 2795.528  & $3p~^2{\rm P}^o_{3/2}$ & $3s~^2{\rm S}_{1/2}$ & 1.3 & 12.3\\
            & 2790.777  & $3d~^2{\rm D}_{3/2}$ & $3p~^2{\rm P}^o_{1/2}$ & 1.9&9.1\\
            & 2797.930  & $3d~^2{\rm D}_{3/2}$ & $3p~^2{\rm P}^o_{3/2}$ & 0.39&4.8\\
            & 2797.998  & $3d~^2{\rm D}_{5/2}$ & $3p~^2{\rm P}^o_{3/2}$ & 3.5&18.3\\
            & \ldots  & $3d~^2{\rm D}_{5/2}$ & $3p~^2{\rm P}^o_{1/2}$ & \ldots&2.5\\
            & \ldots  & $3d~^2{\rm D}_{3/2}$ & $3s~^2{\rm S}_{1/2}$ & \ldots&1.2 \\
            & \ldots  & $3d~^2{\rm D}_{5/2}$ & $3s~^2{\rm S}_{1/2}$ & \ldots&1.8\\
\\
\ion{C}{2}  & 1334.5323  &$2s^22p~^2{\rm D}_{3/2}$ & $2s^22p~^2{\rm P}^o_{1/2}$& 0.26 & 1.4\\
            & 1335.6625  &$2s^22p~^2{\rm D}_{3/2}$ & $2s^22p~^2{\rm P}^o_{3/2}$& 0.051 &0.9\\
            & 1335.7077  &$2s^22p~^2{\rm D}_{5/2}$ & $2s^22p~^2{\rm P}^o_{3/2}$& 0.46 & 2.9\\
            & \ldots     &$2s^22p~^2{\rm D}_{5/2}$ & $2s^22p~^2{\rm P}^o_{1/2}$& \ldots & 0.5\\
\\
\ion{Si}{4} & 1402.77  & $3p~^2{\rm P}^o_{1/2}$ & $3s~^2{\rm P}_{1/2}$ & 0.54 & 7.3\\
\hline
\end{tabular}
\end{center}
{Data are from the NIST spectroscopic database \citep{Kramida+others2015}, wavelengths above 2000
  \AA{} are in air, otherwise they are in vacuum. Collisional data are
from \citet{Sigut+Pradhan1995,Blum+Pradhan1992,Zhang+Sampson+Fontes1990}.
The Maxwellian-averaged collision strengths $\Upsilon(T_e)$ 
are given for $T_e=10^4$ K for the 
singly charged ions, and at $T=10^5$ K for \ion{Si}{4}. 
}
\end{table}}

\shortauthors{D. A. Lacatus, P. Judge, A. Donea}
\shorttitle{rain and flare physics}

\slugcomment{}

\begin{document}

%###############################################################################
%
%     OPENING
%
%###############################################################################

\title{An explanation of remarkable emission line profiles in post-flare coronal rain}
\author{Daniela A. Lacatus} \affil{Center
  for Astrophysics,\\ School of Mathematical Science,
  Monash University,\\ Victoria 3800, Australia;
  \danielaemail}

\author{ Philip G. Judge }
\affil{High Altitude Observatory,\\
       National Center for Atmospheric Research\footnote{The National %
       Center for Atmospheric Research is sponsored by the %
       National Science Foundation},\\
       P.O.~Box 3000, Boulder CO~80307-3000, USA; \philemail}

\and

\author{Alina Donea} \affil{Center
  for Astrophysics,\\ School of Mathematical Science,
  Monash University,\\ Victoria 3800, Australia;
  \alinaemail}

%###############################################################################
%
%     ABSTRACT
%
%###############################################################################

\begin{abstract}
We study broad red-shifted emission in chromospheric and transition
region lines that appears to correspond to a form of post-flare coronal
rain. Profiles of \ion{Mg}{2}, \ion{C}{2} and
\ion{Si}{4} lines were obtained using the \iris{} instrument before, during
and after the X2.1 flare of 11 March 2015 (SOL2015-03-11T16:22). We
analyze the profiles of the five transitions of \ion{Mg}{2} (the
$3p-3s$ $h$ and $k$ transitions, and three lines belonging to the
$3d-3p$ transitions). We use analytical methods to understand the
unusual profiles, together with higher resolution 
observational data of similar phenomena observed 
by {\bf \citet{Jing+others2016}}. 
The peculiar line ratios indicate anisotropic emission from the
strands which have cross-strand line center optical depths ($k$-line)
of between 1 and 10. The lines are broadened by unresolved Alfv\'enic
motions whose energy exceeds the radiation losses in the \ion{Mg}{2}
lines by an order of magnitude. The decay of the line widths 
is accompanied by a decay in the brightness, suggesting a causal
connection.  If the plasma
is $\lta 99$\% ionized,  ion-neutral collisions can account for the dissipation,  otherwise a  of dynamical process seems
necessary.
Our work implies that the
motions are initiated during the impulsive phase, to be 
dissipated as radiation over a period of an hour, predominantly by
strong chromospheric lines. The coronal ``rain'' we observe is far
more turbulent that most earlier reports have indicated, with implications for 
plasma heating mechanisms.
\end{abstract}

\keywords{Sun: atmosphere}

\section{Introduction}

The purpose of the present paper is to analyze some curiously broad,
red-shifted profiles of emission lines obtained with the {\revise Interface Region Imaging Spectrograph
instrument \citep[\iris{}: ][]{dePontieu+others2014}} a few minutes after the
impulsive phase of a flare. The X2.1 flare of 11 March 2015
(SOL2015-03-11T16:22) occurred in the active region (AR) NOAA 12297
and was accompanied by a filament eruption.  Detectable hard X-Ray
emission observed with {\revise Reuven Ramaty High Energy Solar Spectroscopic Imager
\citep[{\em RHESSI}: ][]{Lin+others2002}} began near 16:16
UT, reached a peak at 16:21 UT and returned to its near-quiescent state
near 16:35 UT. The broad, redshifted emission of interest here started
abruptly near 16:26 UT.  
\figmgim
Typical profiles of the \ion{Mg}{2} $h$ and $k$ {\revise (at air wavelengths 
2802.7 \AA\ and 2795.5 \AA, respectively)} lines during this
phase are shown in Figures~\pref{fig:mgim} and
\pref{fig:mgii}. These profiles appear to differ qualitatively from  {\revise all
earlier theoretical and most observational} studies on coronal rain, the emission lines being unusually broad.

By themselves, these data admit several possibilities of
interpretation.  Fortuitously, some partly resolved ``fine structure'' with similar spectral properties {\revise has been 
observed in another 
solar flare \citep{Jing+others2016}.  Their 
work focused on observations of spectral lines with $0\farcs03$ sampling, through
narrow-band filters. Although not discussed by them, Figure 7b in
their paper also shows \ion{Mg}{2} spectral data from \iris{} during the
post-flare stage that, without doubt, arise from the same
phenomenon we analyze here.  Thus, the broad, red-shifted emission
features of Figures~\pref{fig:mgim} and \pref{fig:mgii} 
are coincident with the onset of a return of emitting plasma
along post-flare loops (see  Figures 7c and 7d of {\revise \citealp{Jing+others2016}}).
These phenomena shown by  \citet{Jing+others2016} and in the present work are unquestionably a form of {\em coronal rain}.}

{\revise Usually seen as plasma emission in chromospheric and transition region 
lines (partially ionized or neutral plasma), or as absorption in EUV images, coronal rain forms as a consequence of instability in coronal loops. 
In non-flaring active region loops, thermal instability leads to catastrophic 
cooling \citep{Cally+Robb1991}. Numerous observations and models have studied 
coronal rain \citep[][ and references within]{Kohutova+Verwichte2016,Moschou+other2015}, 
finding typical redshifts of $\sim$80 km s$^{-1}$. These authors have highlighted  multi-stranded and multi-thermal structure, but without reference to broad line widths.
Plasma condensation after eruptive flares can produce similar phenomena \citep{Song+others2016}, although the physical processes may differ.
\citet{Kleint+others2014} have found bursty flows associated with bright-points at the edge of the sunspot umbra, 
that show a striking similarity to the profiles analyzed here, but lasting only $\sim$20 s.}

Here we study the \ion{Mg}{2} and other lines from
\iris{} to try to constrain the physical nature of the {\revise unusually broad lines associated with raining plasma}.  The
various observed parameters, when {\revise judiciously} combined with the results of
\citet{Jing+others2016}, constrain the nature of the raining plasma
and its origins more tightly than has been achieved in the past.  
The primary lines of interest here are listed in Table~{\pref{tab:atomic}}.

\tabletwo

Being a consequence of the unknown mechanisms by which mass and energy
are transported into the solar corona, there are many outstanding
questions concerning coronal rain.  Here we focus upon extracting as
much information from the \iris{} data to study the origins of the
cool material arising from the flare, and its remarkable thermal
structure.  Post-flare chromospheric spectra have been obtained for
decades in lines such as H$\alpha$.  The \ion{Mg}{2} data analyzed
here have advantages over H$\alpha$.  The line 
emissivities and opacities are more simply related to thermal properties of the emitting
plasma (discussed below).  

 We study five \ion{Mg}{2} lines (two of which are blended), collectively
they are sensitive to different thermal conditions. Thus, we can perform a
simple quantitative analysis of the data with minimal assumptions.
Our work complements much recent work
\citep[e.g.][]{Antolin+others2015,Jing+others2016} which studies
broad- or narrow-band imaging, by analyzing the behavior of line
{\em profiles}. 
These profiles differ so dramatically from those previously modeled
using existing numerical methods, that we adopt 
analytical methods.    Our work should
later inspire further numerical simulations, such as that of   
\citet{Fang+Xia+Keppens2013}, which might attempt to better understand the
origin of the enormous linewidths spanning over 300 \velu{}, pointing
to energetic magnetic waves or (less likely) turbulence as the culprit.

\figmgii
\section{Observations}

%\fighmi

 On 11 March 2015, the \iris{} instrument acquired data of 
AR NOAA 12297 during a flare watch campaign.  Data were 
downlinked from nine spectral windows.  Here we
analyze those containing the lines of \ion{Mg}{2}, \ion{C}{2} and \ion{Si}{4}, 
together with data obtained by {\revise the Atmospheric Imaging Assembly \citep[{\em AIA}: ][]{Lemen+others2012} 
and Helioseismic and Magnetic Imager \citep[{\em HMI}: ][]{Schou+others2012}
instruments on the Solar Dynamics Observatory spacecraft \citep[{\em SDO}: ][]{Pesnell+others2012}}. The \iris{} slit was moved 
sequentially and repeatedly to
four positions offset by 0, 2, 4, 6$\arcsec$ in the E-W direction 
on the solar surface. A total of 
1230 such pointings were acquired. We focus on data from exposures 110 to 490.  At each slit position, spectral data
with exposures of 4 seconds was acquired. To complete one cycle of
4 slit positions took {\revise  $\approx 21$ seconds}. During the flare,
the exposure times were decreased for the FUV wavelengths
by an automatic flare-triggered mechanism, to prevent over-exposure,
without affecting the
total raster duration. The projected slit is 0.33$\arcsec$ wide. The
spatial and spectral samplings were 0.33$\arcsec$/pixel and 0.05092
\AA/pixel for the NUV region. For the FUV regions the
spectral sampling {\revise was $0.02544$ \AA/pixel (FUV2) and $0.02596$ \AA/pixel (FUV1)}.

{\revise 
We applied {\tt iris\_orbitvar\_corr\_l2s.pro,} a routine  
in the \iris{} SolarSoft package, to correct for spacecraft orbital variations.}
{\revise Following the procedure in \citet{Liu+others2015}, 
we then obtained wavelength-integrated emission-line intensities in physical units (see Table~2), which 
{\revise refer to the obvious components in emission above the photospheric absorption line and continua.}
 For comparison we 
include the values for a quiet sun (QS) patch outside the 
active region along those from a representative position 
within the peculiar spectrum region at different time instances.
These values are the sum over both line core and extended red-shifted profile, 
including multiple lines where is the case. All velocities are determined relative to the average quiet Sun spectrum.}

IRIS slit-jaw images (SJI) at 1330, 1400 and 2832 \AA{} were also
acquired during the raster steps 0, 3, and 1 respectively.  Each SJI
set had a temporal sampling of 20.76 seconds on average, and spatial
sampling of 0.33$\arcsec$/pixel, spanning a total field of view (FOV)
of $126\arcsec \times 119\arcsec$.  Four  such images are shown
in Figure~\pref{fig:sjib}. 

%\figsji
\figsjib

 The flare appears to have been triggered by the accelerated rise of a
filament starting near 16:13 UT.  Flare emissions started at 16:16
UT, the flaring reaching the emission maxima at 16:20 UT. This maximum
high energy flare emission{\revise , as observed by RHESSI, }was located at (X,Y)=(-354,-171)$\arcsec$,
a region not covered by the spectral slit positions.

\figdopp

We examine the evolution of the spectra at
different locations along the flare ribbon.  We found locations of
extended very broad and redshifted emission at the same locations swept over 
by the evolving and propagating flare
ribbon, lasting for more than an hour after the flaring
event. {\revise To highlight typical Doppler shifts and line widths, in Figure~\pref{fig:dopp} we show the first velocity-weighted moment 
(Doppler shift)  and second moment (line width) of the Mg II k-line, normalized to the zeroth moment\footnote{Moment $i$ is $\int_{\Delta v}\! I_v v^i dv$, where the Doppler shift at wavelength $\lambda$ is $v= c (1-\lambda/\lambda_0)$ and $\lambda_0$ the rest wavelength of the transition.}.  The Figure shows data only for the first slit position. In passing, we note that the smaller amplitude blue-shifts in the umbra probably correspond to umbral flashes, studied for example by \citet{Carlsson+Bard2010}.}

\figmgwidint

{\revise To highlight the evolution of the line width over the whole region of interest, we represented the normalized intensity of the profile as a function of the line width in Figure~\pref{fig:widint}, along with an insert of the same dependence for a quiet sun patch outside the active region. The color represents the time of the exposure, with early times being masked by later emission. The QS distribution is centered around 400 m\AA\ and shows little scatter. The distribution for the peculiar emission region is centered around 500 m\AA\ before the flare and it shows an increase in  both intensity and line width after the flare. The intensity decreases in time to pre-flare values, while the width still has an extended tail up to about 1100 m\AA\ suggesting that the raining phenomenon is still ongoing.}

Figure~\pref{fig:vmgram} shows the position
of the four IRIS slits projected onto an image constructed from 
data from the SDO/HMI instrument. The HMI data are the standard
reductions of the ``720s'' product, including vector
magnetic field data. The field azimuth in the plane
perpendicular to the line of sight (LOS) has not been
disambiguated. 
Note that the four slit positions extend across the entire sunspot
centered near (X,Y)=(-340,-182)$\arcsec$.  The broadened profiles 
are found above the penumbra, they are marked as
the solid components of the dashed lines shown in the figure.  The
lengths and positions of these lines correspond to the average
locations where the anomalously broad profiles occur.  These regions
are quite separate from the magnetic neutral line.  Instead, they lie above
regions of modest LOS field, large perpendicular
field, whose direction follows parallel or anti-parallel to the axis
of a ``dark intrusion'' (Fig.~\pref{fig:vmgram}) centered at
(X,Y)=(-343,-176)$\arcsec$, a location of where systematic upward-directed
photospheric motions of $\approx$1--1.5 km s$^{-1}$ are present (but not shown here).
%While these HMI data are taken after the impulsive phase which was
%near 16:21 UT, those before are very similar.
%\fighmi162400

\figvmgram

Intensity profiles of the \ion{Mg}{2} $h$ and $k$ lines
are shown in Figure \pref{fig:mgii}. Representative data of
interest here are from exposure number 100 in our sample obtained at
16:32:31 UT. In the figure these are shown as dashed lines.  
The time dependence of these profiles is shown in Figure 
\pref{fig:mgim}.   
The \iris{} instrument's  compensation for 
the Sun's average rotation was used, so that the
images show the changing conditions 
over approximately the same areas of the Sun.

For convenience, in 
Figure \pref{fig:mgim}, we identify four episodes in the evolution
of the \ion{Mg}{2}: A pre-flare phase (before 16:20 UT); 
An impulsive phase (16:21 UT); A
relaxation phase (16:21 to 16:27 UT) ; A post-flare (PF) phase
(16:25 UT to beyond 17:00 UT).  
The PF phase is the
subject of the present paper.  These phases are self-evident in the
data. The relaxation phase is simply the transient response
of the line profiles to the 
sudden release of energy in the impulsive phase, judging by the time scale of the 
decay which seems appropriate.  The chromosphere has a thickness of  $\approx1500$ km, and
pressure perturbations (shocks) will propagate a little above the
sound speed of $\approx 7-10$ \velu.  A few sound crossing times corresponds to about 5
minutes.  The relaxation phase is most clearly seen in
the narrow $3d-3p$ lines of \ion{Mg}{2}, close to 2791 and 2798 \AA{}  in
this figure.  The $h$ and $k$ resonance lines of \ion{Mg}{2} are at 2796.3 and 2803.5 \AA{}.
\tableonenew
Profiles of \ion{C}{2} resonance lines near 1335 \AA{} show similar
spectral line shapes to $h$ and $k$, as does the 1403 \AA{} line of
\ion{Si}{4}.  Both lines are slightly broader (in Doppler units) than
the $h$ and $k$ lines.  However, the duration of the broad 
PF profiles in both cases are appreciably shorter. The intensities of these lines
are also 10-100$\times$ smaller than $h$ and $k$
(Table~2). %\pref{tab:intensitiesnew}). 

\section{Analysis}

\subsection{Summary of observed properties}

We focus only on the “postflare” (PF) phase, since these profiles are as
yet unexplained.  The {\revise spatio-temporal behavior of the broad PF profiles
appears to be unrelated to other phases (see Figure~1). They are most simply interpreted as simple (unreversed) emission superposed onto the more``normal" profile represented by exposure 1 shown in Figure 2.} During the PF phase the $h$
and $k$ profiles have several salient features: (1) the 
total intensities of
the broad profiles are larger than the core intensities, and the time variations 
vary largely independently of the core intensities; (2) the PF
$h$ and $k$ intensities are remarkably smooth across the line profile
at each time of observation, but they change a little between
exposures; (3) the intensities are in the ratio $k/h$ = 1.15 across 
almost the entire line profiles (see 
Fig.~\pref{fig:mgii}); (4) the lines are far broader than the net redshift;
(5) the PF emission patches are $\approx4-5$ Mm
across; (6) in both the $h$ and $k$ lines the bright PF profiles begin
abruptly 6-7 minutes after the impulsive phase (see Figure~\pref{fig:mgim} {\revise and \pref{fig:dopp}});
(7) the broad components are not obviously self-reversed; (8) the profiles are very similar across all regions sampled by the spectrometer's slit.

It is interesting to relate these profiles to the other lines.  The PF
$3d-3p$ transitions are far weaker than during the impulsive and
relaxation phases, compared with the $h$ and $k$ lines ($3p-3s)$.  To
the above points we add (9), namely that the $h$ and $k$ lines, and
those of \ion{Si}{4} and \ion{C}{2} all have widths in Doppler
units of about $w=100$ \velu{}, and their centroids are red-shifted by about
+60 \velu{} {\revise (see Figures 4 and 5).}

Point (1) implies that {\em PF profiles are not caused by scattering 
of bright photons from the core}.  Regarding point (3), optically
thin conditions produce ratios of intensities of 2:1.  Hence the PF
plasmas are {\em optically thick in the $h$ and $k$ lines}.
Remarkably, the same is true for the 1334/1335 \ion{C}{2} lines whose
intensities are close to 1:1, compared to the optically thin ratio of
1:2.  Point (2) suggests that the unresolved motions are physically {\em  much} 
smaller than resolvable scales.  Points (4) and (8) suggest that, if 
opacity broadening is negligible, there is {\em more energy in unresolved 
motions (widths) than in resolved motions (shifts)}.  Photon 
scattering can, however, both broaden and shift the lines,
this will be discussed below. The intensities of the various lines at various phases 
are listed in Table~2. 

\subsection{A reference model}

We examine the PF profiles using parameters in a reference model
of the emitting plasma.  We will assume that the plasma exists in
strands with a width {\revise W}$\approx 100$ km \citep{Jing+others2016}, that
has number densities of hydrogen nuclei close to {\revise $10^{12}$}
particles per cm$^{3}$, and that these particles are all at temperatures of $\approx
10^4$ K.  If $T_e$ exceeds $2-3\times 10^4$ K, Mg becomes doubly 
ionized.  The justification for the adopted density is weak. 
{\revise It exceeds that at the very top of the pre-flare
chromosphere by an order of magnitude, but flare models propel mass from deeper layers of the 
chromosphere  higher into the corona, on time scales of a minute or less for M and X class events.
\citep[e.g.][]{Abbett+Hawley1999,Allred+others2005}.  In an X-class
flare (with $\approx 10^{11}$ \fluxu{} of energy flux directed towards
the chromosphere), the mass per unit of ejected material is expected to be 
about
$m = 10^{-2} F_{11}$ g~cm$^{-2}$, estimated from the last panels of 
Figures~3 and 5 of \citet{Allred+others2005}. Here $F_{11}$ is the 
downward-directed energy flux in units of 10$^{11}$ \fluxu.  
If this mass is spread uniformly along a
tube of constant area with length $L$, then the number density on
average is $\approx m/\mu m_H L$, where $\mu=1.36$ is the mean atomic weight of
the  largely neutral pre-flare plasma and $m_H$ is the mass of the hydrogen atom. 
With $L = 10^9 L_{10}$ cm and with $L_{10}$ corresponding to $10$  Mm, we have: 
\be{density}
n_H \approx 5\times10^{12} L^{-1}_{10} F_{11} \ \ {\rm cm^{-3}}
 \ee
This provides a crude justification for our adopted estimate of $n_H=10^{12}$ 
cm$^{-3}$ for the raining plasma after this X2 class flare .}
We use variables normalized to the average values: $W = 10^7 W_{100}$ cm, 
$n = 10^{12} n_{12}$ cm$^{-3}$, and $T = 10^4T_{e4}$ K.  We will also assume that
the unresolved ``microturbulent'' velocities are of order $\xi = 10\xi_{10}$ \velu,
with $\xi_{10}=1$, which is of order the sound speed at $T_{4}=1$. 
In this reference model, $W_{100}=n_{12}=T_{e4}=\xi_{10}$=1.  Note that the observed 
line widths are $w=\xi = 100$ \velu{}, ten times broader than 
the reference width, for reasons to be discussed below\footnote{We will conclude
that $\xi_{10} \approx 10$ will characterize our final choice of best parameters.}. 

The model is used below to explore conditions under which these lines 
might form, understanding that the chosen values are educated guesses. 

\subsection{Radiative transfer}

We consider the formation of the \ion{Mg}{2} lines in a structure that
consists of an array of strands that comprise an episode of coronal
rain.  Thus we look first at the
spectrum emitted by one elemental strand of width $W_{100}=1$.

\subsubsection{Strand optical depths}

Two observations imply that the optical depth of the $k$ line across
the emitting stands is $\gta$ 1, at line center.  Firstly, the
emission is self-excited within the raining plasma itself, i.e. there
is no identifiable dominant external source of irradiation.  Secondly, the $k$/$h$
line ratios are far from the optically thin ratio of 2:1
(Figure~\pref{fig:mgii}).

With an absorption oscillator strength
$f$, the transition between levels 1 and 2 has an opacity 
at line center (in units of cm$^{-1}$) of \citep[e.g.][]{Mihalas1978}
\be{opacity}
\kappa_0 = \frac{\pi e^2}{m_e c} n_1 f / \Delta\nu \ 
        = \  0.0264\  n_1 f/ \Delta\nu,
\ee
\noindent where {\revise$m_e$ is the electron mass, $e$ is the elementary charge,} 
$n_1$ is the number density of \ion{Mg}{2} ions in 
the lower level, and $\Delta\nu$ is the Doppler width of the line in Hz.  
Then, the optical depth across a strand of width $W$ whose axis has an angle 
$\vartheta$ with the line-of-sight is given by: 
\be{mihalas}
\tau_0  = 0.0264\  \frac{n_1 f W}{\mu\,\Delta\nu} , \ \ \mu = \cos \vartheta.
\ee
Let us assume that most Mg atoms in the strands are in the ground
state of \ion{Mg}{2}, then, with a logarithmic abundance of Mg of 7.42 relative to hydrogen 
\citep{Allen1973}, {\revise we have $n_1 = n_H\times 10^{7.42-12}
= 3\times10^6 n_{12}$ cm$^{-3}$.}  With a Doppler width $\xi$ of
10\velu{}, $\Delta\nu \approx 3.6 \times10^{10}\xi_{10}$ Hz, 
and for $k$, $f=0.6$, then
{\revise 
\be{kappascale}
\kappa_0 = 1.1\times10^{-5} n_{12} / \xi_{10}  \ \ {\rm cm^{-1}}
\ee
With
$W_{100}= W/100~{\rm km}$ we have
\be{mihalas2} \tau_0 = 110
\frac{n_{12}}{\xi_{10}} \frac{W_{100}}{ \mu}.
\ee 
%
%Providing that $n_{12}/\mu \xi_{10} \gg 0.1$ then the condition that
%$\tau_0 \gg 1$ will be satisfied.
%
In this reference model the $h$ and $k$ lines have $\tau_0 \approx 10^2$
across the strands.   
}

The observation that the $k:h$ ratio in the PF profile is 1.15, far from the thin ratio 
of 2:1, has two possible explanations.  The first is that the lines 
might be {\em  effectively thick}, a condition that is true for
the \ion{Mg}{2} $h$ and $k$ lines formed in the stratified chromosphere 
\citep[e.g.][]{Linsky+Ayres1978} causing the Sun's $k/h$  
ratio from various non-flaring regions on the 
Sun to be close to 1.1-1.5 \citep[e.g.][]{Kerr+others2015}.  The second possibility is
that the radiation emerging from the strands is anisotropic, and that the $h$ and $k$ lines 
have different anisotropy under conditions of modest optical depth. 

The latter case seems far more likely, as can be seen by comparing the 
escape probability of line photons from the strand core with the 
destruction probability.  In Appendix \pref{app} we show that the
absorption of \ion{Mg}{2} photons by background continuum is
negligible in the reference model. The probability of
destruction by collisions from the $k$ line 
can be evaluated simply as
\be{destruct}
\varepsilon = \frac{C_{31}}{A_{31}+C_{31}}
\ee
where $C_{31}$ is the collision rate from level 3 to level 1, the dominant collision rate out of the line's upper level 
(level 3). {\revise and $A_{31}$ is the Einstein coefficient for spontaneous radiative emission \citep{Allen1973}}.
Using parameters from Table~\pref{tab:atomic} we find 
$A_{31} = 2.7\times 10^8$ sec$^{-1}$, and $C_{31} \approx 
4\times10^4 n_{e12} T_{e4}^{-1/2}$ sec$^{-1}$.   {\revise Here we must use the 
electron density $n_{e12}$ because electrons are responsible for the bulk of the collisions.  
\be{destruct1}
\varepsilon \approx \frac{C_{31}}{A_{31}+C_{31}} \approx 10^{-4} n_{e12}/\sqrt{T_{e4}}.
\ee
}A reasonable
approximation for $P_{esc}$ for resonance lines is 
\be{pesc} P_{esc}
\approx \frac{1 - \exp(-\tau_0)}{\tau_0} 
\ee
\citep[e.g.][]{Frisch1984}, so that in the reference model  for the $k$ line, {\revise using equation (\protect\ref{eq:mihalas2})
then 
\be{escr}
P_{esc} / \varepsilon \approx \frac{10^{2} \xi_{10}\mu  \sqrt{T_{e4}}}{n_{e12}n_{12}W_{100}}.
\ee
}
It seems $P_{esc} \gg \varepsilon$ for reasonable physical parameters.

Together with the observation that the broad profiles are not 
self-reversed, we conclude that the lines are effectively thin
in the coronal rain. 

\subsubsection{$k:h$ ratio and radiation anisotropy}

How can {\em radiation anisotropy} explain the line ratios in the rain
phenomenon?  The lines of $h$ and $k$ differ in opacity by a factor of
two.  We can imagine a simple case of cylindrical strands where the
optical depth of $h$ is $ \approx 1/2$, and that of $k$ is $\approx 1$.
The $h$ line photons can escape directly from deeper within the strand
than the $k$ line photons.  In essence, the $h$ line will radiate more {\revise isotropically}
from the core of the strand, whereas $k$ photons will scatter
preferentially into the path of shortest escape, i.e. perpendicularly
to the main axis of the strand.  The observed 
line ratio will change from $2:1$ even if the total emission into all
directions is in the ratio 2:1.  We would expect $k:h > 2:1$ perpendicular to
the strand's main axis of symmetry, and $k:h < 2:1$ in other directions.  

If the optical depths are substantially higher
($\tau_0(k) \approx 10$, say), then the differential effect is reduced to
the outermost cylinders comprising the strand (assuming it is a cylinder,
cf. \citealp{Lipartito+others2014}), the radiation being more isotropic in the opaque
deeper core of the strand.   

We conclude that we are seeing the strands at a slant angle
far from $90^\circ$ and $0^\circ$ to the strand's main axis, and that 
{\em $\tau_0$ should be $> 1$ but less than $\approx 10$ in the $k$ line,
in these strands.}  {\revise The value of 110 computed from
our reference calculation lies close to
the necessary optical depth, when we note that the nearest boundary 
of the strand for escape is at a distance of at most $W/2$ not $W$, and, as we will argue below,
$\xi_{10}$ is an underestimate, }
thus the core photons will see a boundary $\lta $ 110/2 optical depths distant. 

Related to this problem is the effect of {\em atomic polarization}
when scattering dominates the source function ($\varepsilon \ll 1$).  
The $J=1/2$ upper level of $h$ is not polarizable, and the scattered
intensity of the $J=1/2 \rightarrow J=1/2$ transition ($h$ line) is
isotropic.  However, the polarizable $J=3/2$ upper level's {\em
atomic polarization} can change the observed {\em intensity} ratios under
conditions of anisotropy even
by the ``diffuse'' photons self-generated within the cylinder 
\citep{Landi+Landolfi2004}.  We estimated these effects to be small, 
on the order of a few percent using calculations similar to those 
applied to \ion{He}{1} lines by \citet{Judge+Kleint+Sainz-Dalda2015}.

\subsection{Frequency-integrated intensities}

If a line is {\em effectively thin} across an emitting strand, 
the emission is simply the sum of the
emission along the line-of-sight. To an accuracy of a factor of two, we can ignore
the anisotropy, and we find, for the $k$ line: 
\be{ithin}
I_{thin} \approx 
\frac{h\nu_{31}}{4\pi}{ n_1C_{13}}\frac {W}{\mu} \ \ \tintu.
\ee
{\revise where $C_{13}$ is the collisional excitation rate \citep{Gabriel+Jordan1971,Burgess+Tully1992}. 
Within a factor of 2 or so, we can set $\mu =0.5$, and this expression 
gives, multiplying by two for the sum of the two lines:  
\begin{eqnarray} \nonumber
I_{thin} & \approx &   
\frac{h\nu_{31}}{4\pi}{ n_1C_{13}}\frac {W}{\mu}\\
 &\approx& 1.4\times 10^6 n_{12}n_{e12} \exp \left (4.86(1- \frac{1}{T_{e4}}) \right ) W_{100} \label{eq:int}
\end{eqnarray}
for our reference calculation.   The observed intensities for both lines are 
of the order of $1.6 \times 10^7$ \tintu.   With $T_{e4}=2$ we find $I_{thin} \approx 1.1\times10^7$ \tintu. 
We regard this as reasonable 
agreement with our adopted reference calculation, noting that the calculation depends exponentially
on $T_{e4}$ and on the product $n_{12}n_{e12}$, and allowing for the possibility that there is more than one such
strand of emission along each line of sight observed by {IRIS}.
}
\subsection{Doppler widths and shifts}

The Doppler shifts of 50-60 \velu{} are compatible with the
acceleration of cool plasma elements due to gravity under free fall
from a few Mm above the surface.  But more interesting constraints
come from the observed linewidths.  We work in  units of the 
Doppler broadening parameter, $\xi$.  The
observed lines are $\approx10\times$ broader than our reference model
can produce.  Particle densities are far below those needed for
significant collisional broadening.  An optical depth of $10$ at the
center of the $k$ line would produce photon emission up to about
$\xi_{10} \sqrt{\ln \tau_0} \approx 1.5 \xi_{10}$ from line center,
assuming complete redistribution and using an argument due to
\citet{Osterbrock1962}.

It seems highly unlikely that photon scattering at high optical depths
is the source of the broad line emission.  The lack of a clear self-reversal
near the cores of the PF broad profiles appears to suggest also that
photon scattering is not a major contributor PF profile widths.  We
are forced to conclude that the lines are formed under conditions of
modest optical depth ($\tau_0(k) \lta 10$), but in which the
radiation is anisotropic.

Given our modest estimates of optical depth, the broad emission widths
are some $15-20\times$ the sound speed where \ion{Mg}{2} emission
usually forms ($T_e \lta 2\times 10^4$ K).  Under these extreme conditions it
is likely that shocks would quickly dissipate such small-scale 
motions, if they were purely  hydrodynamic and small-scale.  Thus, 
{\em magnetic fields are responsible for 
the unresolved motions}.

\subsection{\ion{Mg}{2} $h$ and $k$ and the $3d-3p$ transitions}

Unlike the $3p-3s$ transitions, the $3d-3p$ transitions, when in
emission, are certainly not optically thick within the strands. The
lower ($3p$) levels have a population of $\lta \exp (-4.86/T_{e4})$
times the $3s$ lower level of the $h$ and $k$ lines, thus the $3d-3p$
optical depths are orders of magnitude smaller than $h$ and $k$.  The
observed intensity ratios (the blend of the 2797.930 and 2797.998
\AA{} lines, relative to the 2790.930 \AA{} line) are indeed close
to the optically thin ratios of 2:1.

The $3d-3p$ transitions differ from $h$ and $k$ in another essential
way.  Electron excitation to the $3d$ levels from the $3s$ ground
level is optically forbidden.  The cross sections are dominated by core
penetration of the Mg$+$ ion by the incoming electrons.  There is no
long-range interaction.  When the $3p$ level populations are far below
Boltzmann relative to $3s$, as is the case when the lines are
effectively thin (photon escape in $h$ and $k$ reducing the $3p$
populations), collisional excitation of the $3d$ levels occurs
predominantly only from the $3s$ level.  Then, again ignoring
anisotropies in the radiation field, 
\be{ratiothin}
\frac{I_{3d-3p}}{I_{3p-3s}}  \approx  \frac{C_{3s-3d}}{C_{3s-3p} }
\ee
which becomes 
% (3s-3p) 17.9; (3p-3d) 34.7; 3s-3d 3.0
\be{ratiothin1}
\frac{I_{3d-3p}}{I_{3p-3s}}  \approx  \frac{\Upsilon_{3s-3d}}{\Upsilon_{3s-3p} }
\exp\left(-\frac{4.86}{T_{e4}}\right) \approx 
\frac{1}{6}  \exp\left(-\frac{4.86}{T_{e4}}\right)
\ee
Inserting the observed ratio $\approx1/40$ into equation~(\pref{eq:ratiothin1}), 
we find $T_{e4} \approx 2.5$, or $\approx 2.0$ allowing for the radiation anisotropy in the $k$ line. 

We note that the work of \citet{Pereira+others2015} forces the 
formation of the $3d-3p$ transitions into the dense, lower chromosphere,
through their choice of model atmosphere.  In their model, the 
$h$ and $k$ lines are thermalized where the $3d-3p$ transitions form,
and so the $3d$ levels are populated largely through the two allowed
transitions $3s-3p$ followed by $3p-3d$.

\subsection{Dielectronic recombination}

The process of dielectronic recombination (DR) can lead to emission on
the red side of resonance lines \citep[e.g.][]{Gabriel+Jordan1971}. In
the case of \ion{Mg}{2}, $3pnl$ states can form as intermediate states
in a Mg$^+ 3s$ + e$^-$ collision, a process called dielectronic
capture.  If the intermediate states decay radiatively to $3snl$
states emitting a $3p-3s$ photon, called a ``stabilizing transition'',
the process is a dielectronic recombination.  For singly charged ions,
DR rates are typically some $10^{2-3}$ times smaller than the direct
rate for collisional excitation.  For Mg$^+-e^-$ DR the results of
\citet{Altun+others2006} yield a total rate of about 0.003 of the
direct rate.  Therefore, if the broad PF profiles were caused by a
superposition of DR then their intensities would be some $10^{2-3}$
times smaller than that of the core emission. This is not observed
(Figures \pref{fig:mgim} and \pref{fig:mgii}).

\subsection{Sudden onset of post-flare coronal rain emission.}

Inspection of Figure~\pref{fig:mgim} shows that the PF profiles
begin abruptly near 16:26 UT.  {\revise They show a} two-fold rise in
intensity across all wavelengths that takes between 2 and 3 time steps, 
say $t=40$ seconds.  Fast  changes are also seen throughout the PF phase
in  Figure~\pref{fig:mgim}, but with smaller amplitudes.  

These data are difficult to reconcile if we assume that the linewidths
of the emitting plasmas arise from a mixture of LOS velocities on
macroscopic scales.  At a given slit position, various plasma elements
would arrive at different times as they each follow their own
trajectory. Let us suppose that plasma is ejected into the corona from
one or both flare footpoints.  On their way to crossing the slit, they
acquire various LOS velocities as a result of the (unknown) dynamics
in the post-flare tubes of magnetic flux.  Depending on the
trajectories, if one assumes that the PF line profiles are
superpositions of macroscopic flows, then the elements will cross the
slit with a broad distribution of arrival times.  This is contrary to the
observations.  The short rise time $t\approx 40$ seconds across the entire
profile imply that, if caused by
different arrival times, the plasma elements would have to coordinate
themselves such that 50 \velu{} upward moving plasma would ``know''
when to cross the slit as well as the 150 \velu{} downward moving
plasma. Simply put, there would have to be 
a linear relationship between velocity and projected distance $d$
from the slit,
\be{wtf}
 v \cos \vartheta = d \cos \vartheta / t.
\ee
It does not make sense that the Sun would know about the special
position of the \iris{} slit in such a fashion.

What then is the cause of the two-fold rise in intensity across all
wavelengths within $t=40$ seconds?  We can conceive of two
explanations:  (1) The rain plasma has small-scale motions (with a magnitude of $\approx100$
\velu{}) within a coherent large-scale flow, making $\xi_{10} \approx 10$, 
or (2) The plasma is
indeed moving at a variety of macrosopic velocities, but that there is
a special thermodynamic process making the plasma visible on a time scale of
$40$ seconds.  

The second explanation can be discounted because the the history of each 
plasma element determines whether lines of a certain element will become visible
at a certain time.  
For example, if the raining plasma is
unheated and is at densities close to 10$^{11}$ cm$^{-3}$, the cooling
time is of order 100 seconds \citep{Anderson+Athay1989}.  At 100
\velu{} the plasma can travel $\approx 9$ Mm before it cools
significantly.  
Again, we are faced with the untenable position that the \iris{}
slit would have to be a ``special'' place on the Sun for this explanation 
to hold water, each element evolving 
along its own trajectory.  A change in opacity is discounted because we know of no
mechanism to suddenly reveal {\em only} the PF broad profiles at 16:26 UT.

Thus we are left with the first explanation which is similar to real
rain on Earth. 

\subsection{Origins of the broad lines}

The large value of $\xi \approx 100$ \velu{} is still 
consistent with the required optical depths to make the radiation
differentially anisotropic between $h$ and $k$, because for $k$ 
we find a reduction in $\tau_0$ from $13$ to $1.3$, owing to the denominator 
of equation~(\pref{eq:opacity}).
We are then left to
explain why the line width speeds are so high, and the fact that the
line shifts are about half those of the widths.  The red-shifts
correspond to plasma dropped in free fall from a height of $\approx5$ Mm.  It seems
that gravity at least can account for such red-shifts, even if it is
superposed onto more energetic small-scale dynamics. The red-shifts
will therefore concern us no more.  

The field strength needed to achieve an Alfv\'en speed of 100 \velu{}
when $n_{12}=1$ is merely $\approx 17$ G.  This is two orders of
magnitude smaller than the photospheric fields that lie below the
raining plasma region, which have $B \gta 1500$ G
(Figure~\pref{fig:vmgram}).  So it appears that Alfv\'en waves
generated after the impulsive phase can easily carry enough kinetic
energy to account for the broad lines, if otherwise undamped, for some
tens of minutes after the impulsive phase.  The kinetic energy density
of the unresolved motions with $n_{12}=1$ is $\rho \xi^2 \approx $23
erg~cm$^{-3}$.  If these persist along a length $L_{10}=1$ of a
strand, then these motions amount to an energy density per unit area 
of about $2\times
10^{10}$ erg~cm$^{-2}$.  A large X-class flare releases some $10^{11}$
\fluxu{} during the impulsive phase, for some $10^{2-3}$ seconds, 
giving an available energy density per unit area of 
$10^{13-14}$ erg~cm$^{-2}$.   It therefore seems reasonable that 
1 part in 10$^3$ of this impulse might reside in residual wave motions 
in the post-flare magnetic fields, and become evident in the widths of
the lines observed with \iris{} reported here. 
Also, the
strands occupy a small volume of the magnetic structures in which they
exist \citep{Jing+others2016}, therefore considerably more magnetic
energy might feed into the strands from neighbouring plasma not
revealed in the \ion{Mg}{2} or other UV emission lines.

Can MHD {\revise (Magneto-Hydro-Dynamic)} oscillations live for a period of an hour or so during which
the broad \ion{Mg}{2} lines decay?  To answer this question would
require MHD simulations across multiple scales, since the dissipation
of magnetic energy requires the (non-linear) generation of small physical scales.
A recent physical discussion of various incomplete calculations done
to date shows that it may take a very long time to dissipate wave
energy in loops containing small-scale density inhomogeneities
\citep{Cargill+deMoortel+Kiddie2016}, calling into question the
viability at least of phase mixing to describe coronal heating.
Below we consider other explanations.

\section{Discussion}

We have examined all salient features of the peculiar \iris{}
post-flare line profiles of \ion{Mg}{2}, \ion{C}{2} and \ion{Si}{4},
with accompanying slit-jaw and magnetic data.  By elimination we are
led, remarkably, to a consistent picture in which unresolved Alfv\'enic
motions -- waves or turbulence -- are generated and gradually decay
over a period of more than one hour.  The decay of the magnetic
energy is about an order of magnitude larger than the radiation losses
from the strong \ion{Mg}{2} lines, suggesting a causal connection
between them.  The peculiar line ratios and profiles are consistent
with strands of plasma of width 100 km, and they require 
an optical depth across them of between $1$ and $10$ in the $k$ line.

If we apply the observed intensities and ratios  to equations to (\pref{eq:ratiothin1}) and (\pref{eq:int}) to solve for model parameters, 
we have $T_e \approx 2\times10^4$ K, and $n^2_{12} W_{100} \approx 0.4$.  Assuming $W_{100}=1$ then $n_{12} \approx 0.6$. But this analysis 
does not acknowledge that because of the evidence of unresolved motions there is certainly a distribution of plasma with temperature, and that
lines such as the \ion{Mg}{2} $3d-3p$ and \ion{C}{2} 1335 form in hotter plasma than $h$ and $k$.  
Given the exponential dependence of radiative losses on temperature, we therefore suspect, but cannot prove, that for $h$ and $k$ a smaller value of $T_e$ is appropriate.  If this is the case then 
$n_{12} > 0.6$. Below we will adopt $n_{12}=1$ for the sake of argument.  

The total flux radiated during the coronal rain
phenomenon seen in $h$ and $k$ is $\approx \pi I t_r = 5\times 10^7 \times 600 \approx
3\times 10^{10} $ erg~cm$^{-2}$.  Here we have used a relaxation lifetime of 10 minutes appropriate for the first part of the relaxation phase where these line intensities were measured 
(Figure~\pref{fig:mgim}). The {\em total} radiation losses from chromospheric plasma (due to 
losses in calcium, iron, hydrogen,..) are between $4\times$ and $10\times$ larger, with larger values  
occurring at lower plasma temperatures
\citep[][Fig. 4]{Anderson+Athay1989}. We adopt a value of $4\times$ the $h$ and $k$ losses, to arrive at a total time-integrated radiative 
flux of $10^{11}$ erg~cm$^{-2}$.

Above, we found $\rho \xi^2 L \approx
2.3\times 10^{10}n_{12}L_{10}$ erg~cm$^{-2}$.  These rough 
estimates are within a factor of 4 of one another,
so with With $n_{12}L_{10}=4$ we have agreement. 
Perhaps more noteworthy,  
Figure~\pref{fig:mgim} shows that {\em as the broad line 
intensities decrease, so does the power in the fluctuations (linewidths)}.  
It appears that the
Alfv\'enic fluctuations excited during the impulsive phase, might  
decay ultimately into radiation via a process
that slowly (compared with dynamical time scales
$\approx L/V_A$) converts the wave into thermal energy. 

Various dynamical mechanisms 
might achieve this on time scales of an hour, 
such as phase mixing, resonant 
absorption 
\citep[e.g.][]{Narain+Ulmschneider1990, Narain+Ulmschneider1996},
although detailed fully dynamical 
numerical simulations need to be performed to 
address the very long damping times suggested by 
\citet{Cargill+deMoortel+Kiddie2016}.   Gas-kinetic processes can also
efficiently damp oscillations when  
neutral H or He are abundant in the raining plasma. Ion-neutral collisions would 
efficiently damp out high frequency 
wave energy on time scales of 
the ion-neutral collision time,  $\tau_{ni} \approx 0.4n_{n11}/\sqrt{T} $ seconds, 
where $n_{n11}$ is the ambient neutral density in units of 
10$^{11}$ particles cm$^{-3}$  \citep[eq. 16,][]{Holzer+Flaa+Leer1983}. 
At wave frequencies $\omega$ below the inverse of the collision time, 
equation~17 of \citet{Holzer+Flaa+Leer1983} applies, 
which gives time scales of 
\be{in}
2/ (\tau_{ni} \omega^2),  \ \ \omega \tau_{ni}  \ll 1.
\ee
For example, a three-minute period Alfv\'en wave has $\omega \approx
1/50$ rad~sec$^{-1}$, $\omega \tau_{ni} \approx 10^{-4}$ and then
the damping time becomes $20/n_{n11}$ seconds.  If the plasma were 1\% neutral 
then the observed decay time of a few thousand seconds 
would be naturally explained without needing to invoke
dynamical MHD processes. 

Lastly, the \iris{} profiles resemble those seen in flares of very
active stars \citep{Linsky+others1989}.  It is likely our solar
analysis can help better understand such enormous flares.

\bibliographystyle{aasjournal}
\bibliography{biblio}  

\appendix
\section{Background opacities near 2800 \AA{}}
\label{app}

For absorption in the background continuum with opacity $\kappa_C$,
effectively thick conditions prevail when 
\be{back}
P_{esc} < \kappa_C/\kappa_0.
\ee
At 2800 \AA{}, the dominant source of continuous opacity within the
strands is probably the Balmer continuum of hydrogen.  This can be
computed approximately assuming that the $n=2$ levels of hydrogen are
populated not too far from LTE relative to the proton density $n_p$, 
because the radiation temperature in the Balmer continuum 
($\approx 6000$K) and electron temperature are within a factor of two
or so.  Then 
\be{kappac}
\kappa_C =  n_2 \sigma \approx \frac{n^*_2}{n^*_p}n_p \sigma 
\ee
Let $n_e$ be the electron density in cm$^{-3}$, $T_e$ the electron
temperature in K. Then 
with \citep{Mihalas1978}
\be{lte}
\frac{n^*_2}{n^*_p} = 2.07\times10^{-16} n_e \frac{g_2}{g_p} T_e^{-3/2} 
\exp\left(\frac{I_2}{kT_e}\right), 
\ee
using $I_2 = 3.399$ eV, and using scaled values 
$n_{e12} = n_e / 10^{12}$ cm$^{-3}$, $T_{e4} = T_e/10^4$ K, and with 
$\sigma \approx 10^{-17}$ cm$^{2}$ \citep{Allen1973}, then 
\be{kappacn} \kappa_C \approx 1.7\times 10^{-14} n_{e12} T_{e4}^{-3/2}
\ \exp \left(3.945/T_{e4}\right) n_{p12} \ee Then we find, further
assuming $n_p \approx n_e$, and dropping subscripts {\em p,e} for
convenience, \be{ratio} \kappa_C/\kappa_0 \approx 2.5\times10^{-9}
n_{12} \exp \left(3.945/T_{4}\right)\xi_{10} / T_4^{3/2} \ee Given
equation ~(\pref{eq:mihalas2}), this ratio would need to be $\approx
0.01$ for the line to be effectively thick due to background continuum
absorption.  If we set extreme conditions (photosphere-like) of
$n_{12} \approx 10^1$, $T_4 \approx 0.5$, we can obtain a ratio of
$7\times10^{-5}$.  

Therefore we can safely ignore continuum absorption at 2800 \AA{} 
in the coronal rain.  

\end{document}